\documentclass[aps,superscriptaddress,a4paper,12pt]{revtex4}
\usepackage[colorlinks=true, pdfstartview=FitV, linkcolor=blue, citecolor=red, urlcolor=black]{hyperref}
\usepackage[latin1]{inputenc}
\usepackage{amsmath}
\usepackage{amsfonts}
\usepackage{amssymb}
\usepackage{graphicx}
\usepackage{comment}	

\newcommand{\be}{\begin{equation}}
\newcommand{\ee}{\end{equation}}
\newcommand{\ben}{\begin{eqnarray}}
\newcommand{\een}{\end{eqnarray}}
\newcommand{\bes}{\begin{subequations}}
\newcommand{\ees}{\end{subequations}}
\def\bal#1\eal{\begin{align}#1\end{align}}

\newcommand{\wt}{\widetilde}

\newcommand{\bfi}{\begin{figure}}
\newcommand{\efi}{\end{figure}}
\newcommand{\bc}{\begin{center}}
\newcommand{\ec}{\end{center}}

\newcommand{\sech}{\mbox{sech}}

\newcommand{\LL}{{\cal L}}
\newcommand{\LX}{{\cal L}_X}
\newcommand{\LXX}{{\cal L}_{XX}}

\newcommand{\LXs}{{\cal L}_{X_s}}
\newcommand{\LXXs}{{\cal L}_{X_sX_s}}
\newcommand{\LXsp}{{\cal L}_{X_s\phi}}

\begin{document}

\title{Stability of kinklike structures in generalized models}
\author{I. Andrade}\email{andradesigor0@gmail.com}
\affiliation{Departamento de F\'\i sica, Universidade Federal da Para\'\i ba, 58051-970 Jo\~ao Pessoa, PB, Brazil}
\author{M.A. Marques}\email{mam.matheus@gmail.com}\affiliation{Departamento de F\'\i sica, Universidade Federal da Para\'\i ba, 58051-970 Jo\~ao Pessoa, PB, Brazil}
\author{R. Menezes}\email{rmenezes@dce.ufpb.br}
\affiliation{Departamento de Ci\^encias Exatas, Universidade Federal
da Para\'{\i}ba, 58297-000 Rio Tinto, PB, Brazil}
\affiliation{Departamento de F\'\i sica, Universidade Federal da Para\'\i ba, 58051-970 Jo\~ao Pessoa, PB, Brazil}

\begin{abstract}
We study the stability of topological structures in generalized models with a single real scalar field. We show that it is driven by a Sturm-Liouville equation and investigate the conditions that lead to the existence of explicit supersymmetric operators that factorize the stability equation and allow us to construct partner potentials. In this context, we discuss the property of shape invariance as a possible manner to calculate the discrete states and their respective eigenvalues.
\end{abstract}

\maketitle

\section{Introduction}
Scalar fields are the simplest ones in field theory and are useful to study topological structures, in particular, kinks, which arise in $(1,1)$ spacetime dimensions \cite{manton, vachaspati}. Usually, they are studied in an action whose associated Lagrangian density is given by the difference between a dynamical term and a potential term. This model is the standard one and engender solutions that minimizes the energy of the system \cite{bogopaper,pspaper}. Their stability under small fluctuations is investigated through an equation of the Schr\"odinger type whose zero mode always exists and is related to the presence of a translational invariance in the model.

The standard Lagrangian, however, is not the only manner to seek for models that engender topological configurations. This can be seen in Refs.~\cite{sen1,sen2}, where the singular tachyon kink emerged in the context of strings and branes. In particular, in Ref.~\cite{sen1}, it was studied the so-called singular tachyon kink in a field theory that describes the dynamics of a D-brane in the context of superstring theory. Moreover, in Ref.~\cite{babichev1}, it was introduced classes of non-canonical models that support topological structures, in which the potential is unchanged and the term that controls the dynamics of the field in the Lagrangian density is generalized.

Later, in Ref.~\cite{genkink}, the authors investigated generalized models as an arbitrary functions of the field and its standard dynamical term; see also Ref.~\cite{zhong1}. Despite the complications introduced by the generalizations, it is possible to find a first order formalism that is compatible with the equations that minimizes the action and govern the non-canonical system. Nevertheless, the generalized nature of the models makes a Sturm-Liouville eigenvalue equation \cite{slref} arise in the study of the linear stability, instead of the Schr\"odinger-like one as in the standard scenario. Usually, the Sturm-Liouville equation is harder to deal with because it has terms of first order. A possible manner to overcome this difficulty is to make a change of variables to transform this equation into a Schr\"odinger-like one. Notwithstanding that, this change cannot always be done with analytical expressions, since one has to integrate and invert functions involved in the process. So, in general, one cannot avoid to investigate the stability through the Sturm-Liouville equation.

A motivation to study non-canonical models comes from the cosmology, in the context of inflation \cite{ginf}. They present distinct features when compared to the standard ones. For instance, in this case, there may be no need of a potential to drive the inflation. These models were also used as a tentative solutions to the cosmic coincidence problem, i.e., to explain why the universe is expanding at a late stage of its evolution \cite{ccoinc1,ccoinc2}. In field theory, generalized models are also useful in the study of twinlike models, which engender the same topological solutions and their respective energy densities \cite{twin1,twin2,twin3}.

The presence of the aforementioned difficulties in the study of the stability of generalized models motivated us to develop a deep investigation of this issue. In this paper, we deal with the properties of the Sturm-Liouville equation that arises with generalized scalar field models, such as the zero mode and hyperbolicity. We also develop a procedure to factorize the operator associated to the stability in supersymmetric partners. As in the standard case, we investigate some models whose stability leads to shape invariant potentials. So, in this process, in the context of Sturm-Liouville, we also make an inspection in the property of shape invariance, which is a condition for exact solvability that is useful to construct the modes and their respective eigenvalues associated to the stability equation.

The paper is organized as follows: in Sec.~\ref{sec2}, we discuss the general features of our model, focusing on the stability of the static one-dimensional topological solutions under small fluctuations. We also show how the Sturm-Liouville eigenvalue equation that drives the small fluctuations can be written in terms of supersymmetric operators. The subsection \ref{sip} is dedicated to the shape invariance property. In Sec.~\ref{sec3}, we present specific models to illustrate how our procedure works in some examples. First, we review the standard case. Then, we investigate the method in two types of generalized models in subsections~\ref{1ex} and \ref{2ex}. In Sec.~\ref{conclusions} we present our conclusions and perspectives for future works.

\section{Generalities}\label{sec2}

We consider the action of a single real scalar field, $\phi$, in a two-dimensional flat spacetime with metric tensor $\eta_{\mu\nu} = \textrm{diag}(+,-)$:
\be\label{action}
{\cal S}=\int d^2x \LL(\phi,X),
\ee
where $X=\frac12\partial_\mu\phi\partial^\mu\phi$ denotes the standard dynamical term of $\phi$. We remark that the standard case is given by $\LL = X - V(\phi)$, with $V(\phi)$ denoting the potential. One can vary this action with respect to the field to get the equation of motion
\be\label{eom}
\partial_\mu(\LX\partial^\mu\phi)=\LL_\phi,
\ee
where $\LX=\partial\LL/\partial X$ and $\LL_\phi=\partial\LL/\partial \phi$. We expand it to get
\be\label{eome}
\LL_{X\phi}\partial^\mu\phi\partial_\mu\phi + \LXX\partial^\mu\phi\partial^\nu\phi\partial_\mu\partial_\nu\phi + \LX\Box\phi = \LL_\phi.
\ee
Invariance over spacetime translactions, $x^\mu \to x^\mu + a^\mu$, with $a^\mu$ constant, leads to 
the energy-momentum tensor
\be\label{tmunu}
T_{\mu\nu} = \LX\partial_\mu\phi\partial_\nu\phi - \eta_{\mu\nu}\LL.
\ee
The components are given explicitly by
\bes
\begin{align}
T_{00} &= \LX {\dot{\phi}}^2 - \LL, \\
T_{01} &= T_{10} = \LX \dot{\phi}\phi^\prime, \\
T_{11} &= \LX {\phi^\prime}^2 +\LL.
\end{align}
\ees
In the above equations, the dot and the prime denote the derivatives with respect to $t$ and $x$, respectively. Furthermore, we define the energy density as $\rho\equiv T_{00}$ and the stress as $\sigma \equiv T_{11}$. Since we are dealing with a very general model, we let ourselves be guided by the null energy condition (NEC); that is, we impose $T_{\mu\nu}n^\mu n^\nu \geq 0$, where $n^\mu$ is a null vector, obeying $\eta_{\mu\nu}n^\mu n^\nu = 0$. This condition restricts the model in a manner that the Lagrangian density must obey the inequality
\be\label{nec}
\LX \geq 0,
\ee
for a general $\phi(x,t)$.

We consider the static configurations. In this case, the equation of motion \eqref{eome} becomes
\be\label{seom}
\left(2\LXXs X_s + \LXs\right)\phi'' = 2\LXsp X_s -\LL_\phi,
\ee
where $X_s = -\phi'^2/2$, with the $s$ index denoting static configurations. Since we are interested in topological solutions, we use the boundary conditions $\phi(\pm\infty)\to v_\pm$, where $v_\pm$ are constants that represent the asymptotic values of the field. The non zero components of the energy-momentum tensor in Eq.~\eqref{tmunu} give the following expressions for the energy density and stress
\bes\label{stmunu}
\begin{align}\label{rho}
\rho &= - \LL_s, \\
\sigma &= \LXs {\phi^\prime}^2 +\LL_s.
\end{align}
\ees
One can proceed as in Ref.~\cite{genkink} and perform a rescale in the solution to show that the stability against contractions and dilations is satisfied by the stressless condition, $\sigma=0$, which leads to
\be\label{seomd}
\LL_s - 2\LXs X_s = 0.
\ee

Let us now focus on the the linear stability of the solutions. We introduce time-dependent small fluctuations, $\eta(x,t)$, a real function, around the static solution, $\phi(x)$, in the form $\phi(x,t)=\phi(x)+\eta(x,t)$. Considering up to first-order in contributions of $\eta$, we can write
\be
X = X_s + \partial_\mu\phi\partial^\mu\eta.
\ee
This modifies the following terms as
\be
\LL_\phi \to \LL_\phi + \LL_{\phi\phi}\eta + \LL_{\phi X_s}\partial_\mu\phi\partial^\mu\eta \quad\text{and}\quad \LX \rightarrow \LL_{X_s} + \LL_{X_s \phi}\eta + \LXXs\partial_\mu\phi\partial^\mu\eta.
\ee
Substituting the field $\phi(x,t)$ in the equation of motion \eqref{eom}, we then get
\be
\LL_{X_s}\ddot{\eta}-\left[(2\LL_{X_sX_s}X_s+\LL_{X_s})\eta'\right]' = \left[\LL_{\phi\phi}+\left(\LL_{\phi X_s}\phi'\right)'\right]\eta .
\ee
The above equation allows us to separate time and space in the fluctuations with the expression $\eta(x,t)=\sum_n\eta_n(x)\cos(\omega_nt)$. By doing so, we obtain a Sturm-Liouville eigenvalue equation
\be\label{stabomega}
-\left[(2\LL_{X_sX_s}X_s+\LL_{X_s})\eta_n'\right]' = \left(\LL_{\phi\phi}+\left(\LL_{\phi X_s}\phi'\right)'+\omega_n^2\LL_{X_s}\right)\eta_n .
\ee
This equation can be expanded to
\be
-(2\LL_{X_sX_s}X_s+\LL_{X_s})\eta_n'' - (2\LL_{X_sX_s}X_s+\LL_{X_s})'\eta_n' = \left(\LL_{\phi\phi}+\left(\LL_{\phi X_s}\phi'\right)'+\omega_n^2\LL_{X_s}\right)\eta_n.
\ee
Notice the above equation contains a term of first derivative in $\eta_n$ and the weight function $\LL_{X_s}$ with the eigenvalue $\omega_n^2$ that does not appear on the Schr\"odinger equation. To preserve the hyperbolicity of Eq.~\eqref{stabomega}, we define the quantity $A$ and impose the condition 
\be\label{hyper}
A^2\equiv\frac{(2\LL_{X_sX_s}X_s+\LL_{X_s})}{\LL_{X_s}}> 0.
\ee
Since we are working with generalized models, the inner product of the eigenfunctions $\eta_n$ presents a non negative weight function where the solution exists. In this case, the weight is given by $\LL_{X_s}$, whose non negativity is ensured by the NEC in Eq.~\eqref{nec}. We then write the orthonormality condition
\be
\int^\infty_{-\infty} \eta_m(x) \eta_n(x) \LXs dx = \delta_{mn}.
\ee
An important issue is that the zero mode $\eta_0$, which is the mode that we name for $\omega_0=0$, always exists, even in the Sturm-Liouville scenario. We can see this by observing the equation that describes it, which comes from \eqref{stabomega} with $n=0$:
\be
-\left[(2\LL_{X_sX_s}X_s+\LL_{X_s})\eta_0'\right]' = \left(\LL_{\phi\phi}+\left(\LL_{\phi X_s}\phi'\right)'\right)\eta_0.
\ee
To find the form of $\eta_0$, one must take the spatial derivative in both sides of Eq.~\eqref{seom} and compare the result with the above equation. This procedure allows one to show that the zero mode is related to the derivative of the static solution such that
\be
\eta_0=\kappa\phi^\prime,
\ee
where $\kappa$ is a normalization constant. Notice this result does not depend on the specific form of the Lagrangian density. The static solution $\phi(x)$ is stable if $\omega_n^2\geq0, \,\forall n$. This means that stable solutions engenders the zero mode as their state with lowest eigenvalue.

We can make a deeper analysis of the stability equation \eqref{stabomega}, which may be rearranged into
\be
-\frac{1}{\LL_{X_s}}\left(\frac{d}{dx}A^2\LL_{X_s}\,\frac{d}{dx} + \left(\LL_{\phi\phi}+\left(\LL_{\phi X_s}\phi'\right)'\right)\right)\eta_n = \omega_n^2\eta_n.
\ee
The left side is a differential operator. So, we can write this equation in the form
\be\label{sleq}
L\eta_n = \omega_n^2\eta_n.
\ee
Here, $L$ is the Sturm-Liouville operator, given by
\be\label{sturmop}
L = -\frac{1}{\LL_{X_s}}\frac{d}{dx}A^2\LL_{X_s}\frac{d}{dx}+U(x),
\ee
where the stability potential, $U(x)$, is written as
\be
U(x) = -\frac{1}{\LXs}\left(\LL_{\phi\phi}+{(\LL_{\phi X_s}\phi')}^\prime\right).
\ee
One can show the operator in Eq.~\eqref{sturmop} is self-adjoint, obeying the expression
\be
\int^\infty_{-\infty} \eta_m(x) L\,\eta_n(x) \LL_{X_s}dx = \int^\infty_{-\infty} \left(L^\dagger\eta_m(x)\right)^\dagger\eta_n(x) \LL_{X_s}dx,
\ee
with the fluctuations satisfying the boundary condition $\left.A^2 \LL_{X_s}\left[\eta_m(x){\eta_n}^\prime(x)-{\eta_m}^\prime(x)\eta_n(x)\right]\right|^\infty_{-\infty} = 0$, which appears from a surface term after integrating the left hand of the above equation by parts. We now try to factorize the operator $L$ in Eq.~\eqref{sturmop} in terms of the following supersymmetric operators $S$ and $S^\dagger$:
\be\label{sopm}
S = A\left(-\cfrac{d}{dx} +M(x)\right) \quad\text{and}\quad S^\dagger = A\left(\cfrac{d}{dx} +M(x) +\frac{\left(A\LL_{X_s}\right)^\prime}{A\LL_{X_s}}\right).
\ee
To do so, we have to impose another condition: $\left.A \LL_{X_s}\left[\eta_m(x){\eta_n}(x)\right]\right|^\infty_{-\infty} = 0$. The above operators lead to
\be\label{slsup}
L_1=S^\dagger S = -\frac{1}{\LXs}\frac{d}{dx}A^2\LXs\frac{d}{dx}+U_1(x),
\ee
where
\be\label{u1}
U_1(x) = A^2 M^2+\cfrac{(A^2\LXs M)^\prime}{\LXs}.
\ee
So, we need to find the function $M(x)$ that satisfies $L_1=L$, with $U_1(x) = U(x)$. In other words, $S^\dagger S\eta_n=\omega_n^2\eta_n$, which is the same of
\be\label{estmetan}
-\left(A^2\LXs\eta_n'\right)' +\left((A^2\LXs M)'+A^2\LXs M^2\right)\eta_n = \omega_n^2\LXs\eta_n,
\ee
and must reproduce the stability equation \eqref{sleq}. 

Following the supersymmetric theory of quantum mechanics, the supersymmetric partner associated to the Sturm-liouville operator in Eq.~\eqref{slsup} can also be calculated; it has the form
\be\label{slsuppart}
L_2=SS^\dagger = -\frac{1}{\LXs}\frac{d}{dx}A^2\LXs\frac{d}{dx}+U_2(x),
\ee
where
\be\label{u2}
	U_2(x) = A M\left(AM +\frac{(A\LXs)^\prime}{\LXs}\right)-A\left(AM +\frac{(A\LXs)^\prime}{\LXs}\right)^\prime.
\ee
In this scenario, the aforementioned potentials $U_1(x)$ and $U_2(x)$ are called supersymmetric partner potentials.

Since the function $M(x)$ does note depend on the states $\eta_n$, we can use the zero mode in the Eq.~\eqref{estmetan} and in Eq.~\eqref{sleq} to get
\be
\left(A^2\LXs\phi''\right)' =\left((A^2\LXs M)'+A^2\LXs M^2\right)\phi'.
\ee
This equation is satisfied by $M=\phi^{\prime\prime}/\phi^{\prime}$, which makes the operators in Eq.~\eqref{sopm} being written by
\be\label{sop}
S = A\left(-\frac{d}{dx} +\frac{\phi^{\prime\prime}}{\phi^\prime}\right) \quad\text{and}\quad S^\dagger = A\left(\frac{d}{dx} + \frac{\phi^{\prime\prime}}{\phi^\prime} +\frac{\left(A\LL_{X_s}\right)^\prime}{A\LL_{X_s}}\right).
\ee
Hence, we now have a supersymmetric factorization for the Sturm-Liouville operator in Eq.~\eqref{sturmop}. In this case, the potentials in Eqs.~\eqref{u1} and \eqref{u2} are written as
\begin{subequations}\label{u1u2}
\bal
U_1(x) &= A^2\left(\frac{\phi^{\prime\prime\prime}}{\phi^\prime} +\left(\frac{\LXs^\prime}{\LXs} +2\frac{A^\prime}{A}\right)\frac{\phi^{\prime\prime}}{\phi^\prime}\right),\\
	U_2(x) &= A^2\Bigg(\left(\frac{\LXs^\prime}{\LXs} +2\frac{\phi^{\prime\prime}}{\phi^\prime}\right)\frac{\phi^{\prime\prime}}{\phi^\prime} -\frac{\phi^{\prime\prime\prime}}{\phi^\prime}\nonumber -\frac{1}{A}\left(\cfrac{(A\LXs)^\prime}{\LXs}\right)^\prime\Bigg).
\eal
\end{subequations}
The above partner potentials are associated to the study of the Sturm-Liouville equation \eqref{sleq}. In some cases, they may engender the so-called shape invariance property, which we investigate below.

\subsection{Shape Invariance}\label{sip}
The supersymmetric quantum mechanics associated to the Sturm-Liouville equation that arise from the field theory described by the action in Eq.~\eqref{action} has the two partner potentials in Eq.~\eqref{u1u2}. It is possible due to the existence of the function $M(x)$ that allows for the factorization of the Sturm-Liouville operator. An interesting fact is that, in some specific cases, these potentials support the shape invariance property, which we investigate here.

By considering the operators in Eq.~\eqref{slsup}, we get the eigenvalue equation $L_1 \eta^{(1)}_n = \big(\omega^{(1)}_n\big)^2\eta^{(1)}_n$; here, we are using the superscript $(1)$ to denote the eigenfunctions and eigenvalues associated to this equation. On the other hand, we also have an equation for the supersymmetric partner \eqref{slsuppart}, in the form $L_2 \eta^{(2)}_n = \big(\omega^{(2)}_n\big)^2\eta^{(2)}_n$; the superscript $(2)$ represent the eigenfunctions and eigenvalues associated to this partner equation. Even though we are dealing with Sturm-Liouville operators, one can follow Ref.~\cite{sip2} to show that the partner eigenstates and eigenvalues are related in a similar form of the Sch\"odinger-like case, as
\bes
\begin{align}
\omega^{(2)}_{n} &= \omega^{(1)}_{n+1}, & \omega^{(1)}_{0}&=0,\\
\eta^{(2)}_n &= S\eta^{(1)}_{n+1}/\omega^{(1)}_{n+1}, & \eta^{(1)}_{n+1} &= S^{\dagger}\eta^{(2)}_n/\omega^{(2)}_{n}.
\end{align}
\ees
For convenience, we define $S_{\{1\}}\equiv S$, whose associated Sturm-Liouville operator is $L_{\{1\}}=S_{\{1\}}^\dagger S_{\{1\}}$, which reproduces the stability equation \eqref{sleq}, and the partner $L_{\{2\}} = S_{\{1\}}S^\dagger_{\{1\}}$. We may write the operator $L_{\{2\}}$ in terms of new operators $S_{\{2\}}$ and $S^\dagger_{\{2\}}$ as $L_{\{2\}}=S^\dagger_{\{2\}}S_{\{2\}}+\big(\omega^{(1)}_1\big)^2$, where the lowest eigenvalue is $\big(\omega^{(2)}_0\big)^2 = \big(\omega^{(1)}_1\big)^2$. Following these lines, we can generate a third Sturm-Liouville operador $L_{\{3\}} = S_{\{2\}}S^\dagger_{\{2\}}+\big(\omega^{(1)}_1\big)^2$, and use it to construct the operators $S_{\{3\}}$ and $S^\dagger_{\{3\}}$, as $L_{\{3\}} = S^\dagger_{\{3\}}S_{\{3\}}+\big(\omega^{(2)}_1\big)^2$. This can be done recursively, such that we can build multiple operators $L_m$ with eigenvalues and eigenstates respectively given by
\be
\omega^{(m)}_{n} = \omega^{(m-1)}_{n+1} =...= \omega^{(1)}_{n+m-1} \quad\text{and}\quad \eta^{(m)}_n \propto S_{\{m-1\}}...S_{\{1\}}\eta^{(1)}_{n+m-1}.
\ee

In general, the aforementioned function $M(x)$, depends on a real parameter $a$, so we denote it by $M(x;a)$, which is used to build the partner operators $S(a)$ and $S^\dagger(a)$. The shape invariance property is a condition for exact solvability since it relates these operators and allows for the construction of the eigenstates and eigenvalues; see Refs.~\cite{sip0,sip1,sip2}. We may write it as
\be
S(a_1)S^{\dagger}(a_1) = S^{\dagger}(a_2)S(a_2) +R(a_1).
\ee
Here, $a_2=f(a_1)$, with $f$ being an arbitrary function and $R(a_1)$ is a non null remainder independent of $x$. This expression can be used to show that $U_{1,2}(x;a)$ are Shape Invariant Potentials (SIP) if the condition
\be\label{SIPgen}
U_2(x;a_1) = U_1(x;a_2) +R(a_1)
\ee
is satisfied. From the above expression, we see $R(a_1)$ represents the spacing between the ground states of the two partner potentials. After successive applications of this method, we can determine algebraically that
\be\label{omegaetasip}
 \big(\omega^{(1)}_n\big)^2 = \sum_{k=1}^{n} R(a_k) \quad\text{and}\quad \eta^{(1)}_n(x;a_1) \propto \prod_{k=1}^{n} S^\dagger(a_k)\eta^{(1)}_0(x;a_{n+1}).
\ee
where the parameter $a_k$ is given by the application of $f$ in $a_1$ successively by $k$ times, that is,
\be\label{aksip}
a_k=f(\underbrace{\ldots}_{k\, \textrm{times}} f(a_1)).
\ee

In this paper, we make use of a simpler form of the shape invariance property, in which the parameters $a_1$ and $a_2$ are related through a shift. So, for simplicity, we call $a_1=a$ and $a_2=a -\lambda$, where $\lambda$ is a real parameter. In this case, Eq.~\eqref{SIPgen} can be written as
\be\label{SIP}
U_2(x;a) = U_1(x;a-\lambda) + R(a)
\ee
and Eq.~\eqref{aksip} as $a_k=a-(k-1)\lambda$.

We then work out the shape invariance associated to the case in which $A$ is constant and one has $M(x;a)=-a\tanh(\lambda x)$, which is controlled by the parameters $a$, and the weight function has the form $\LXs={\sech}^{\frac{b}{\lambda}}(\lambda x)$, where $b$ is real a parameter that controls it. The above function allows for the construction of the supersymmetric operators $S$ and $S^\dagger$
\be\label{ssip}
S(a) = -A\left(\frac{d}{dx} +a\tanh(\lambda x)\right) \quad\text{and}\quad S^\dagger(a) = A\left(\frac{d}{dx} -(a+b)\tanh(\lambda x)\right),
\ee
and the supersymmetric partner potentials
\bes
\bal
U_1(x;a) &= A^2\left[a(a+b) -a(a+b+\lambda)\,\sech^2(\lambda x)\right],\\
U_2(x;a) &= A^2\left[a(a+b) -(a-\lambda)(a+b)\,\sech^2(\lambda x)\right],
\eal
\ees
which are compatible with the condition in Eq.~\eqref{SIP} for shape invariance, with remainder $R(a) = A^2\lambda(2a+b-\lambda)$. Since we are dealing with SIP, the expressions in Eq.~\eqref{omegaetasip} are useful to calculate the eigenstates $\eta_n$ and their respective eigenvalues $ \big(\omega^{(1)}_n\big)^2$ for the potential $U_1(x;a)$. They are given by
\be\label{etaomegasip}
\eta^{(1)}_n(x) = \sech^{\frac{a}{\lambda}-n}(\lambda x)P_n^{(s-n,s-n)}(\tanh(\lambda x)) \quad\text{and}\quad
\big(\omega^{(1)}_n\big)^2 = A^2\lambda^2 n(2s-n),
\ee
where $s=(2a+b)/2\lambda$ and $P_z^{(l,m)}$ denotes the Jacobi Polynomials of argument $z$ and parameters $l$ and $m$. In this case, $n=0,1,\ldots,\lceil{s-1}\rceil$, where $\lceil{z}\rceil$ denotes the ceiling function with argument $z$.

\section{Models}\label{sec3}
In this section, we present some specific generalized models that falls in the class of Lagrangian densities that we consider in Eq.~\eqref{action}. However, before do so we review the standard model, which is described the simplest Lagrangian density, given by
\be\label{lstd}
\LL = X -V(\phi),
\ee
where $V(\phi)$ denotes the potential that must present two neighbor minima. Notice that the NEC in Eq.~\eqref{nec} is always satisfied since $\LX=1$ here. The static configurations must obey Eq.~\eqref{seom}, which leads to $\phi^{\prime\prime} = V_\phi$. To ensure the stability under rescaling, we also impose the stressless condition; in this case, the field must obey the first order equation \eqref{seomd}, which leads to $-X_s=V(\phi)$, or
\be
\frac12{\phi^\prime}^2 = V(\phi).
\ee
The boundary conditions for the field here are related to the potential; the solutions must connect two adjacent minima of $V(\phi)$. A well know model is the $\phi^4$, whose associated potential is $V(\phi) = (1-\phi^2)^2/2$ and the kink profile is described by $\phi=\tanh(x)$. One may use Eq.~\eqref{rho} to show that the energy density is $\rho(x)=\sech^4(x)$, which may be integrated to lead to energy $E=4/3$. Another known model comes from a non polynomial potential, the sine-Gordon one, given by $V(\phi) = \cos^2(\phi)/2$. Since this potential is $\pi$ periodic, it supports a set of minima that are located at $\phi_k=(k-1/2)\pi$ and maxima
at $\phi_m=k\pi$ with $k\in\mathbb{Z}$. The central sector is defined by the minima $\phi=\pm\pi/2$, where the kink solution $\phi=\arcsin(\tanh(x))$ lives. To find the other sectors, one can make the shift $\phi\to \phi+k\pi$. The energy density of the solution in any sector is calculated from Eq.~\eqref{rho}, which leads to $\rho(x)=\sech^2(x)$ and energy $E=2$.
The stability for a general potential in the standard case is described by Eq.~\eqref{stabomega}, which becomes
\be
-\eta_n^{\prime\prime} + U(x)\eta_n = \omega_n^2\eta_n, \quad\text{with}\quad U(x)=V_{\phi\phi}.
\ee
The hyperbolicity condition in Eq.~\eqref{hyper} is always satisfied, because $A^2=1$ here. Furthermore, we see the Sturm-Liouville equation simplifies to a Schr\"odinger-like equation whose associated supersymmetric operators are $S=-d/dx + \phi^{\prime\prime}/\phi^\prime$ and $S^\dagger = d/dx + \phi^{\prime\prime}/\phi^\prime$. In particular, for the aforementioned $\phi^4$ potential, we have the operators $S=-d/dx -2\tanh(x)$ and $S^\dagger = d/dx -2\tanh(x)$, and the modified P\"oschl-Teller stability potential $U(x) = 4-6\,\sech^2(x)$, whose eigenvalues for the discrete states are $\omega^2 = 0$ and $3$. In the case of sine-Gordon potential, the supersymmetric operators are $S=-d/dx -\tanh(x)$ and $S^\dagger = d/dx -\tanh(x)$, and the stability potential is $U(x)=1-2\,\sech^2(x)$, which is of the same type for the $\phi^4$ case with different coefficients, and admits only the zero mode, $\omega^2=0$.

Below, we illustrate our procedure with two generalized models that support solutions of the kink type.

\subsection{First example}\label{1ex}
First, we make a generalization of the standard model in Eq.~\eqref{lstd}, by taking higher powers on the dynamical term. We take the Lagrangian density
\be\label{lmodel1}
\LL = \frac{X}{N}|2X|^{N-1}-V(\phi),
\ee
where $N>1/2$ is a real number that controls the kinetic term of the scalar field, with $V(\phi)$ being potential, as usual. We avoid the case $N=1/2$ because, as we will see, this case is associated to the cuscuton term \cite{cuscuton,cuscuton2,cuscuton3} which does not contribute to the first order equation and this would make the potential being null. The NEC in Eq.~\eqref{nec} reads $|2X|^{N-1}\geq0$, which is satisfied for any $N$, in particular for the range that we consider here. It is straightforward to see that the standard case in Eq.~\eqref{lstd} is recovered by taking $N=1$.

We then consider static configurations and use Eq.~\eqref{seom} to show that the equation of motion is
\be\label{seom1}
(2N-1){\phi^\prime}^{2(N-1)}\phi^{\prime\prime} = V_\phi.
\ee
As in the standard case, $\phi(x)$ must connect two neighbor minima of the potential. The non vanishing components of the energy-momentum tensor are given by the Eq.~\eqref{stmunu}, which reads
\bes\label{stmunu1}
\begin{align}
\label{rho1}\rho &= \frac{1}{2N}{\phi'}^{2N} + V(\phi),\\
\sigma &= \frac{2N-1}{2N}{\phi'}^{2N} - V(\phi).
\end{align}
\ees
As we have shown in the previous section, the stability under rescaling requires the stressless condition, $\sigma=0$. This leads to the first order equation
\be\label{seomd1}
\frac{2N-1}{2N}{\phi'}^{2N} = V(\phi).
\ee
As we have commented before, for $N=1/2$ we would get null potential, which does not give rise to any interesting field configurations in this scenario. The above equation relates the derivative of the field and the potential. This feature allows the energy density in Eq.~\eqref{rho1} to be written as
\be
\begin{split}
	\rho &= {\phi'}^{2N} = \frac{2N}{2N-1}V(\phi).
\end{split}
\ee

We now focus on the linear stability for this specific model, driven by the Lagrangian density in Eq.~\eqref{lmodel1}. From Eq.~\eqref{stabomega}, we get the Sturm-Liouville equation
\be\label{stabmonom}
-\frac{A^2}{{\phi'}^{2(N-1)}}\left({\phi'}^{2(N-1)}\eta_n'\right)^{\prime} + U(x)\eta_n = \omega_n^2\eta_n,
\ee
where the stability potential $U(x)$ is given by
\be\label{umonom}
	U(x) = \frac{A^2}{2N-1}\frac{V_{\phi\phi}}{{\phi'}^{2(N-1)}} = A^2\!\left(\!\frac{\phi^{\prime\prime\prime}}{\phi^\prime} + 2(N-1)\frac{{\phi^{\prime\prime}}^2}{{\phi^\prime}^2} \!\right)
\ee
and $A^2$ is as defined in Eq.~\eqref{hyper},
\be\label{amonom}
A^2 = 2N-1,
\ee
which does not depend on $x$. Notice that, even for the generalization in the Lagrangian density in Eq.~\eqref{lmodel1}, the hyperbolic condition is still obeyed, since $A^2$ is always positive for $N>1/2$, as we have taken in our model. As one knows, the above equation may be written in terms of the operator $L$, as in Eqs.~\eqref{sleq} and \eqref{sturmop}. 

As we have discussed in the previous section, the Sturm-Liouville operator may be factorized as $L=S^\dagger S$, with the operators $S$ and $S^\dagger$ given by Eq.~\eqref{sop}, which reads
\be\label{smonomial}
	S = A\left(-\frac{d}{dx} + \frac{\phi^{\prime\prime}}{\phi^\prime} \right) \quad\text{and}\quad 
	S^{\dagger} = A\left(\frac{d}{dx}+(2N-1) \frac{\phi^{\prime\prime}}{\phi^\prime}\right).
\ee
We can clearly see that the above equation has a structure that is similar to the one for the standard case, $N=1$, except for the presence of constant factors of $A$ in the operators. One can show the Sturm-Liouville operator $L=S^\dagger S$ in Eq.~\eqref{slsup}, which is associated to the stability equation \eqref{stabmonom}, is
\be
L=S^\dagger S = -\frac{A^2}{{\phi'}^{2(N-1)}}\frac{d}{dx}\left({\phi'}^{2(N-1)} \frac{d}{dx}\right) +U_1(x),
\ee
where $U_1(x)=U(x)$ is the stability potential described by Eq.~\eqref{umonom}. We may also find the supersymmetric partner, $SS^\dagger$, in Eq.~\eqref{slsuppart},
\be
S S^\dagger= -\frac{A^2}{{\phi'}^{2(N-1)}}\frac{d}{dx}\left({\phi'}^{2(N-1)}\frac{d}{dx}\right) +U_2(x),
\ee
in which the partner potential, $U_2(x)$, is
\be
U_2(x) = -(2N-1)^2\left(\frac{{\phi^{\prime\prime\prime}}}{{\phi^\prime}} -2\frac{{\phi^{\prime\prime}}^2}{{\phi^\prime}^2}\right).
\ee
If possible, one may consider the supersymmetric partner potentials to use shape invariance and calculate the modes and their corresponding eigenvalues associated to the Sturm-Liouville stability potential.

Considering that kinklike structures usually decay with exponential tails, i.e., $\phi(x)\mp v_\pm\propto\exp(-|x|)$ for $x\to\pm\infty$, we can estimate how the eigenstates $\eta_n$ behave in this regime. In this case, since $M=\phi^{\prime\prime}/\phi^\prime$, we can conclude that it tends to constant values as $x$ goes far away from the origin; we take $M|_{x\to\pm\infty} = M_\pm$. Since $A^2=2N-1$ is constant, one can see that both operators in Eq.~\eqref{smonomial} become a derivative plus a constant term. By using this behavior in the stability equation \eqref{stabmonom}, one can show it becomes
\be
-\eta_\pm^{\prime\prime}-2(N-1)M_\pm\eta_\pm^{\prime}+A^2M_\pm\eta_\pm = \frac{\omega^2}{A^2}\eta_\pm
\ee
Its solution describes the general asymptotic behavior of the fluctuations, which is given by
\be\label{etafreemonom}
\eta_\pm = \exp\left( -(N-1)M_\pm x + ikx\right),
\ee
where $k=\sqrt{\omega^2/A^2 -N^2M^2_\pm}$.

The above expression depends on the sign of $M_\pm$, which obeys $\mp M_\pm>0$ for kinks. For $\omega^2>A^2N^2M_\pm^2$, $k$ is a real number, so the oscillations are present and we have a continuum of states. In this situation, the fluctuations vanish asymptotically for $1/2<N<1$, oscillate all over the space for $N=1$ and explode for $N>1$. Other possibility appears when $k=0$, in which we also get continuum states and the fluctuations tends to vanish at infinity for $1/2<N<1$, is constant for $N=1$ and diverges for $N>1$. Finally, if $\omega^2<A^2N^2M_\pm^2$, $k$ is a purely imaginary number in a manner that its term, and also the entire argument of the exponential, becomes real. In this case, the fluctuations may vanish, be constant or explode asymptotically, depending on the sign of the expression $-(N-1)M_\pm - |k|$; the states are discrete.

Next, we illustrate our procedure with some examples that lies in the monomial class described by the Lagrangian density in Eq.~\eqref{lmodel1}. As our first example in the class \eqref{lmodel1}, we consider a generalization of the aforementioned $\phi^4$ potential, in the form
\be\label{pot1}
V(\phi)=\frac{2N-1}{2N}(1-\phi^2)^{2N},
\ee
with $N$ is the same parameter that controls the scalar field dynamics in the Lagrangian density in Eq.~\eqref{lmodel1}. In this case, the field profile that arises from the first order equation \eqref{seomd1} is described by
\be\label{sol1}
\phi(x)=\tanh(x),
\ee
which is the same solution of the $\phi^4$ model with standard Lagrangian density. Even though the modification that we introduced with the parameter $N$ leaves the solution untouched, the energy density in Eq.~\eqref{rho1} depends on $N$, with the form
\be
\rho(x)=\sech^{4N}(x).
\ee
One can integrate the above expression all over the space to show that the energy is given by
$E=2^{4N-1}B(2N,2N)$, where $B(z,\tilde{z})$ denotes the Beta function with arguments $z$ and $\tilde{z}$.

We then must investigate if the modifications introduced by the parameter $N$ in the model destabilizes the solution in Eq.~\eqref{sol1}. To do so, we follow the formalism in the previous section. In this case, one can calculate the supersymmetric operators $S$ and $S^\dagger$ in Eq.~\eqref{smonomial} that factorizes the stability equation \eqref{stabmonom}; they are written below.
\be\label{smonomtanh}
S = -A\left(\frac{d}{dx} +2\tanh(x)\right) \quad\text{and}\quad
S^\dagger = A\left(\frac{d}{dx} -2(2N-1)\tanh(x)\right).
\ee
These operators are well defined all over the real line. This ensures the linear stability of the model. We may go even deeper and take a closer look into the stability potential in Eq.~\eqref{umonom}, which becomes
\be\label{umonom1}
U(x) = A^2\left(4(2N-1)-2(4N-1)\,\sech^2(x)\right).
\ee
This potential can be associated with the shape invariance property. So, we may use some of the results obtained in Sec.~\ref{sip} with Eqs.~\eqref{ssip}-\eqref{etaomegasip} for $a=2$, $b=4(N-1)$ and $\lambda=1$ to calculate all the discrete states
\be\label{etaphi4like}
\eta_n(x)=\sech^{2-n}(x)P_n^{(2N-n,2N-n)}(\tanh(x)),
\ee
whose associated eigenvalues are $\omega_n^2 = A^2n\left(4N-n\right)$ for $n=0,1,2,\ldots,\lceil{2N-1}\rceil$. We can use Eq.~\eqref{etafreemonom} to calculate the asymptotic behavior of the continuum states that arise when $\omega^2>4A^2N^2$. We have
\be
\eta_\pm = \exp\left( \pm2(N-1)x + ikx\right),
\ee
where $k=\sqrt{\omega^2/A^2-4N^2}$.

We continue the illustration of the formalism by taking the potential
\be\label{pot2}
V(\phi)=\frac{2N-1}{2N}\cos^{2N}(\phi).
\ee
This potential engenders the very same structure of minima of the aforementioned sine-Gordon potential, which is recovered here for $N=1$. The central sector is defined by the interval $[-\pi/2,\pi/2]$ and the other sectors can be found through the shift $\phi\to \phi+ k\pi$, with $k\in\mathbb{Z}$. In this case, the solution obtained from Eq.~\eqref{seom1} is
\be\label{sol2}
\phi(x)=\arcsin(\tanh(x)).
\ee
Similarly to the previous example, it does not depend on $N$. Nevertheless, the energy density in Eq.~\eqref{rho1} becomes
\be
\rho(x)=\sech^{2N}(x).
\ee
By integrating this expression all over the space, we get the energy $E=2^{2N-1}B(N,N)$. So, $N$ controls the energy of the sine-Gordon solution. We conduct the stability analysis as before; one can show the operators $S$ and $S^\dagger$ in Eq.~\eqref{smonomial} are given by
\be
S = -A\left(\frac{d}{dx} +\tanh(x)\right) \quad\text{and}\quad
S^\dagger = A\left(\frac{d}{dx} -(2N-1)\tanh(x)\right).
\ee
Since the above operators do not have divergences, they ensure the linear stability of our solution. The stability potential in Eq.~\eqref{umonom} is written as
\be
U(x) = A^2\left(2N-1-2N\,\sech^2(x)\right).
\ee
As in the previous example, we can take advantage of the shape invariance associated to this potential. We use the results in in Sec.~\ref{sip} with Eqs.~\eqref{ssip}-\eqref{etaomegasip} for $a=1$, $b=2(N-1)$ and $\lambda=1$ to see that 
\be\label{etasglike}
\eta_n(x)=\sech^{1-n}(x)P_n^{(N-n,N-n)}(\tanh(x)),
\ee
with eigenvalues $\omega_n^2 = A^2n\left(2N-n\right)$ for $n=0,1,2,\ldots,\lceil{N-1}\rceil$. For $\omega^2>A^2N^2$, we get continuum states, whose asymptotic behavior is described by Eq.~\eqref{etafreemonom}
\be
\eta_\pm = \exp\left( \pm(N-1)x + ikx\right),
\ee
where $k=\sqrt{\omega^2/A^2-N^2}$.

\subsubsection{Presence of the cuscuton term}
The generalized model described by the Lagrangian density in Eq.~\eqref{lmodel1} leads to a constant $A$, as one can see in Eq.~\eqref{amonom}. This makes the transformation to a Schr\"odinger-like equation through a change of variables become possible to be done with analytical functions; see Refs.~\cite{genkink}. Nevertheless, there are models that do not allow us to perform the aforementioned change, since there is an integration and an inversion of function involved in the process, which are not always feasible analytically. Here, we study the addition of the cuscuton term \cite{cuscuton,cuscuton2,cuscuton3} in the Lagrangian density \eqref{lmodel1}. We then consider
\be\label{lcusc}
\LL = \frac{X}{N}|2X|^{N-1} + f(\phi)\frac{2X}{\sqrt{|2X|}}-V(\phi),
\ee
where $f(\phi)$ is an arbitrary function that drives the cuscuton term. As we have previously shown in Ref.~\cite{cuscuton3}, this function cannot be eliminated through a field redefinition due to the presence of the monomial dynamics. In this case, the cuscuton term does not contribute to the equation of motion for static fields, which is1 given by Eq.~\eqref{seom1}, with $\phi(x)$ connecting two adjacent minima of the potential.

The nonvanishing components of the energy-momentum tensor are given by Eq.~\eqref{stmunu}, which reads
\bes\label{stmunucusc}
\begin{align}
\label{rhocusc}\rho &= \frac{1}{2N}{\phi'}^{2N} + f(\phi)\,|\phi^\prime| + V(\phi),\\
\sigma &= \frac{2N-1}{2N}{\phi'}^{2N} - V(\phi).
\end{align}
\ees
Notice that the cuscuton term does not contribute to the stress of the solutions. Therefore, the stressless condition leads to the Eq.~\eqref{seomd1}; this means that the cuscuton term does not change the kink profile. Nevertheless, as we can see in the above equation, it modifies the energy density of the model.

The stability of the static solutions under small fluctuations is described by Eq.~\eqref{sleq}, whose hyperbolicity is controlled by the function $A$ in Eq.~\eqref{hyper}. We then get

\be\label{a2cusc}
A^2= \frac{(2N-1){\phi^\prime}^{2N-1}}{{\phi^\prime}^{2N-1}+f(\phi)}.
\ee
By setting $f(\phi)=0$ one recovers the function in Eq.~\eqref{amonom}. One may consider the change of variables in Ref.~\cite{genkink} to get a Schr\"odinger-like equation, which is related to the form of $A$. The presence of the function $f(\phi)$, however, brings nonlinearities the problem that complicates this process, since $A$ must used in an integration at some point. Therefore, the investigation of the Sturm-Liouville equation \eqref{sleq} is very important in this case. We then factorize the stability equation with the supersymmetric operators in Eq.~\eqref{sop}. If these operators are well-defined, the model with the presence of the cuscuton term in Eq.~\eqref{lcusc} is stable under small fluctuations. As an illustration, we consider the potential in Eq.~\eqref{pot1} and
\be\label{fcusc}
f(\phi) = \alpha\left(1-\phi^2\right)^p.
\ee
The kink solution is given by Eq.~\eqref{sol1}, the same for the case $\alpha=0$. The energy density in this case has the form
\be
\rho(x) = \sech^{4N}(x) + \alpha\,\sech^{2p+2}(x).
\ee
It can be integrated all over the space so we can obtain the energy $E=2^{4N-1}\,B(2N,2N) + 2^{2p+1}\alpha B(p+1,p+1)$. The function that controls the hyperbolicity in Eq.~\eqref{a2cusc} is written as
\be
A^2 = \frac{2N-1}{1+\alpha\,\sech^{2(p+1-2N)}(x)},
\ee
which is not constant as before.

The stability equation is given by Eq.~\eqref{stabmonom} with the stability potential \eqref{umonom} and $A$ given by Eq.~\eqref{a2cusc}. For our example, in particular, $U(x)$ is described by Eq.~\eqref{umonom1}, with the above non-constant function $A$. This equation can be factorized with the supersymmetric operators in Eq.~\eqref{sop}. The operator $S$ is as in Eq.~\eqref{smonomtanh}, but its supersymmetric partner is cumbersome, so we omit it here. Even so, both operators are regular all over the space; this ensures the stability of our model.

We may go further and investigate the general asymptotic behavior of the fluctuations $\eta(x)$. We consider $M|_{x\to\pm\infty}=M_\pm$, $A|_{x\to\pm\infty}=A_\pm$ and $\ln(A\LXs)^\prime|_{x\to\pm\infty}=K_\pm$ in Eq.~\eqref{sleq} for $x\to\pm\infty$, which becomes
\be
-\eta^{\prime\prime}-K_\pm\eta^{\prime}+\left(M_\pm^2 +M_\pm K_\pm\right)\eta = \frac{\omega^2}{A_\pm^2}\eta.
\ee
The above equation admits the following solution:
\be
\eta_\pm = \exp\left( -\frac{K_\pm x}{2} + ikx\right),
\ee
with $k=\sqrt{\omega^2/{A^2_\pm}- (2M_\pm + K_\pm)^2/4}$. We then consider the solution in Eq.~\eqref{sol1} and $f(\phi)$ given by Eq.~\eqref{fcusc} for $p>2N-1$. In this case, we have $M_\pm=\mp 2$ and $A_\pm\to\sqrt{2N-1}$, such that
\be
	\ln(A\LXs)'=-\tanh(x)\Bigg(\!4(N-1)+\frac{\alpha\,(p+1-2N)}{\alpha+\cosh^{2(p+1-2N)}(x)}\!\Bigg),
\ee
in a manner that it tends to $K_\pm\to\mp 4(N-1)$, asymptotically. We also have $k=\sqrt{\omega^2/(2N-1) -4N^2}$, so the states with $\omega^2>4N^2(2N-1)$ cannot be normalized.

\subsection{Second example}\label{2ex}
We now consider another class of models, described by the Lagrangian density with the form
\be\label{lmodel3}
\LL = V(\phi)F(X),
\ee
where $F(X)$ and $V(\phi)$ are in principle arbitrary functions of $X$ and $\phi$, respectively. The general properties of the above Lagrangian density were investigated in Ref.~\cite{genkink}. One may expand this Lagrangian density up to first order in $X$, around $X=0$, to get
\be
\LL = V(\phi) \left(F(0) + F_X(0)\,X\right) +V(\phi) \mathcal{O}\!\left(X^2\right).
\ee
By making the change $\phi=h(\chi)$, where $h(\chi)$ is the solution of the differential equation $F_X(0)V(h(\chi))\,h^2_\chi = 1$, the standard Lagrangian density can be found
\be\label{lstdredef}
\LL \approx Y -\wt{V}(\chi),
\ee
where $Y=\frac12 \partial_\mu\chi\partial^\mu\chi$ and $\wt{V}(\chi) =-F(0)\,V(h(\chi))$. To comply with this, we consider functions $F(\chi)$ that obey $F(0)<0$ and $F_X(0)>0$.

The equation of motion \eqref{seom} for static configurations is written as
\be\label{seom3}
(2F_{X_sX_s}X_s+F_{X_s})\phi^{\prime\prime} = \frac{V_\phi}{V}\left(2F_{X_s} X_s-F(X_s)\right).
\ee
In this case, the nonvanishing components of the energy-momentum tensor Eq.~\eqref{stmunu} are
\bes\label{stmunu3}
\begin{align}
\label{rho3}\rho &= -F(X_s)V(\phi),\\
\sigma &= V(\phi)\left(F(X_s)-2F_{X_s} X_s\right).
\end{align}
\ees
As we have shown in Sec.~\ref{sec2}, stability under contractions and dilations leads to the stressless condition, $\sigma=0$, which is described by
\be\label{seomd3V}
V(\phi) (F(X_s)-2F_{X_s} X_s)=0,
\ee
This equation may admit two types of solutions, depending on the explicit form of $F(X)$. The simplest case arises for $X_s$ constant in the algebraic equation
\be\label{seomd3}
F(X_s)-2F_{X_s} X_s=0,
\ee
One must be careful when choosing $F(X)$, because it has to allow for the presence of negative $X_s$ solutions in Eq.~\eqref{seomd3}. To do so, one must take $F(X)$ such that $F(X_s)F_{X_s}<0$, i.e., $F(X_s)$ and $F_{X_s}$ have opposite signs. If this condition is satisfied, the solution has the form
\be\label{sol3}
\phi(x)=\alpha x,
\ee
where $\alpha$ is a real constant. In this case, $X_s=-\alpha^2/2$ is constant, as well as $F(X)$ and all its derivatives when evaluated at the above solution. It is clear that the solutions do not depend on the form of the potential. The above solution, for instance, can be obtained for $F(X) = -\exp\left(-X^2/\alpha^2\right)$. However, the potential plays an important role in the model since it controls the energy density in Eq.~\eqref{rho3}.

The second possibility for $X_s$ in the equation \eqref{seomd3V} appeared in Refs.~\cite{sen1,sen2} . It arises for $F(X)=-\sqrt{1-2X}$, which leads to the so-called singular tachyon kink
\be\label{solsen}
\phi(x)=
\begin{cases}
-\infty,\,\,\,& x < 0,\\
0, \,\,\, & x=0,\\
\infty,\,\,\,& x > 0.
\end{cases}
\ee
We remark that the above solution may also appear for other functions $F(X)$. For instance, one may take $F(X) = (4X-1)/\sqrt{1-2X}$ to show that this exotic solution satisfies Eq.~\eqref{seomd3V}.

At this point, we study the stability of the above solutions. Using Eq.~\eqref{stabomega} and the equation of motion \eqref{seom3}, we can write
\be\label{stabtach}
-\frac{1}{F_{X_s}\,V}\left(A^2 F_X\,V\,\eta^\prime\right)^\prime + U(x)\eta = \omega^2\eta,
\ee
with the stability potential given by
\be\label{utach}
U(x) = \frac{F(X_s)-2F_{X_s} X_s}{F_{X_s}\,V}\left(\frac{V^2_\phi}{V} -V_{\phi\phi}\right)
\ee
and the hyperbolicity being controlled by
\be\label{hyper3}
A^2 =1+ \frac{2F_{X_sX_s}X_s}{F_{X_s}}.
\ee
The expressions in Eqs.~\eqref{stabtach}-\eqref{hyper3} are valid for both the solutions in Eqs.~\eqref{sol3} and \eqref{solsen}. For the simplest case, with the solution in Eq.~\eqref{sol3}, we have constant $X_s$, which makes $A^2$ become constant that must be positive to obey the condition $A^2>0$. In this case, the stability equation \eqref{stabtach} takes a simpler form
\be\label{stab3}
-\frac{A^2}{V}\left(V\eta_n^\prime\right)^\prime = \omega_n^2\eta_n.
\ee
The Sturm-Liouville operator $L$ in Eq.~\eqref{sturmop} can be factorized as
$L=S^{\dagger}S$, with
\be\label{soptach}
S = -A\,\frac{d}{dx} \quad\text{and}\quad S^{\dagger} = A\left(\frac{d}{dx}+\phi'\frac{V_\phi}{V}\right).
\ee
Notice the operator $S$ does not have the function $M$ that appears in Eq.~\eqref{sop} is not present here, because of the form of the solution in Eq.~\eqref{sol3}. Notice the stability potential in Eq.~\eqref{utach} is null for the solution in Eq.~\eqref{sol3} and does not appear in Eq.~\eqref{stab3}. The partner operator, $L_2=SS^\dagger$, takes the form
\be
SS^\dagger=-\frac{A^2}{V}(V\eta_n') +  U_2(x)\eta_n,
\ee
where $U_2(x)=-{\phi}^\prime A^2(V_\phi/V)^\prime$.
 
Considering the solution in Eq.~\eqref{sol3}, we can illustrate this model with the potential
\be\label{vgauss}
V(\phi)=e^{-\lambda\phi^2}.
\ee
We comment that, in this case, the field redefinition that leads to the standard Lagrangian in Eq.~\eqref{lstdredef} cannot be done through the use of analytical expressions. 

The Eq.~\eqref{stab3} becomes
\be\label{stabgauss}
-A^2\eta_n^{\prime\prime} +2A^2\alpha^2\lambda x\,\eta_n^\prime = \omega_n^2\eta_n.
\ee
This equation controls the stability of our model with potential in Eq.~\eqref{vgauss}. In this case, the operators in Eq.~\eqref{soptach} become
 The operators
\be
S = -A\frac{d}{dx} \quad\text{and}\quad S^{\dagger} = A\left(\frac{d}{dx} -2\alpha^2\lambda x\right).
\ee
These operators a regular and well defined all over the space. This ensures the stability of our model. We may go further and investigate the states and their eigenvalues. In this case, we have $U_1(x)=0$ and $U_2(x)=2A^2\alpha^2\lambda$. So, both potentials are constant. In order this case, we do not have shape invariant potentials. Nevertheless, the stability equation \eqref{stabgauss} is known since a similar version appears in the study of the Harmonic Oscilator. So, one can calculate the eigenstates and eigenvalues in the form
\be
\eta_n(x)=\kappa_n H_n(\alpha\sqrt{\lambda} x) \quad\text{and}
\quad \omega_n^2=2nA^2\alpha^2\lambda,
\ee
where $n$ is a natural number, $H_n(z)$ denotes Hermite polynomials of argument $z$ and $\kappa_n$ is a normalization constant. Making use of the orthonormality condition for our system, $\int^\infty_{-\infty} \eta_m(x) \eta_n(x) \LL_{X_s}dx = \delta_{mn}$, one can calculate the normalization constant, which is given by $\kappa^2_n=\alpha\sqrt{\lambda}/\left(\sqrt{\pi}2^n n!\right)$.

\section{Conclusions}\label{conclusions}
In this paper, we have studied some aspects of the stability of kinklike structures in generalized scalar field models. We have reviewed the basic properties of the non-canonical model described by the action in Eq.~\eqref{action}, such as the equation of motion, the stressless condition and the energy density. Then, we have investigated the stability of the static solutions under small fluctuations, which is controlled by a Sturm-Liouville eigenvalue equation. As we have commented, one may try to change the variables as it was done in Ref.~\cite{genkink} in order to get a Sch\"odinger-like equation, but this is a hard task that is not always feasible analytically because it involves integrations and inversions of functions in the process. So, understanding the properties of the Sturm-Liouville equation is important.

We have shown the stability equation may be associated to an operator which comes with a potential. An interesting result is that, despite the generalized form of the Lagrangian density, we always can factorize it in supersymmetric operators that can be written explicitly in terms of derivatives of the static solution. So, in this sense, we have found a connection between generalized scalar field models and supersymmetric Sturm-Liouville theory. The presence of the aforementioned supersymmetric operators gives rise to a partner potential. In this context, we have investigated the property of shape invariance, which is useful to calculate the general form of the eigenstates and eigenvalues associated to our stability equation.

As perspectives, one may consider to investigate the stability of generalized models with several scalar fields in the lines of Ref.~\cite{genkink2} or with the inclusion of terms with higher derivatives of the field as in Refs.~\cite{gal1,gal2}, to seek for the conditions that allows for the factorization of the stability equations that arise in the problem. Another possibility is to search for supersymmetric extensions of our action in Eq.~\eqref{action}, following the direction of Refs.~\cite{susy1,susy2,susy3,susy4}, and verify, among other properties, which term makes the coupling between the fields and how the zero mode is calculated for the fermionic sector of the model.

\acknowledgments{We thank Dionisio Bazeia for the discussions that have contributed to this work. We would like to acknowledge the Brazilian agencies CNPq and CAPES for partial financial support. IA thanks support from CNPq grant 140490/2018-3, MAM thanks support from CAPES grant 88887.463746/2019-00 and RM thanks support from CNPq grant 306504/2018-9.}

\end{document}